\documentclass[aps,prl,reprint,preprintnumbers,showpacs,showkeys,superscriptaddress,nofootinbib,floatfix]{revtex4-1}
\usepackage{amsmath,amssymb,bm,mathrsfs}
\usepackage{times}
\usepackage[colorlinks=true,citecolor=blue,linkcolor=blue]{hyperref}

\begin{document}
\title{Sneutrino Chaotic Inflation and Landscape}

\author{Hitoshi Murayama}
\email{hitoshi@berkeley.edu, hitoshi.murayama@ipmu.jp}
\affiliation{Department of Physics, University of California,
  Berkeley, California 94720, USA}
\affiliation{Theoretical Physics Group, Lawrence Berkeley National
  Laboratory, Berkeley, California 94720, USA}
\affiliation{Kavli Institute for the Physics and Mathematics of the
  Universe (WPI), Todai Institutes for Advanced Study, University of Tokyo,
  Kashiwa 277-8583, Japan}

\author{Kazunori Nakayama}
\email{kazunori@hep-th.phys.s.u-tokyo.ac.jp}
\affiliation{Department of Physics, University of Tokyo, Tokyo 113-0033, Japan}
\affiliation{Kavli Institute for the Physics and Mathematics of the
  Universe (WPI), Todai Institutes for Advanced Study, University of Tokyo,
  Kashiwa 277-8583, Japan}

\author{Fuminobu Takahashi}
\email{fumi@tuhep.phys.tohoku.ac.jp}
\affiliation{Department of Physics, Tohoku University, Sendai 980-8578, Japan}
\affiliation{Kavli Institute for the Physics and Mathematics of the
  Universe (WPI), Todai Institutes for Advanced Study, University of Tokyo,
  Kashiwa 277-8583, Japan}

\author{Tsutomu T. Yanagida}
\email{tsutomu.tyanagida@ipmu.jp}
\affiliation{Kavli Institute for the Physics and Mathematics of the
  Universe (WPI), Todai Institutes for Advanced Study, University of Tokyo,
  Kashiwa 277-8583, Japan}

\begin{abstract}
  The most naive interpretation of the BICEP2 data is the chaotic
  inflation by an inflaton with a quadratic potential.  When combined
  with supersymmetry, we argue that the inflaton plays the role of
  right-handed scalar neutrino based on rather general considerations.
  The framework suggests that the right-handed sneutrino tunneled from
  a false vacuum in a landscape to our vacuum with a small negative
  curvature and suppressed scalar perturbations at large scales. 
\end{abstract}
\preprint{UCB-PTH-14/08, IPMU14-0096, TU-966, UT-14-19}
\maketitle

\paragraph{Introduction}
---Discoveries of the $B$-mode polarization by BICEP2
\cite{Ade:2014xna} and the Higgs boson by ATLAS \cite{Aad:2012tfa} and
CMS \cite{Chatrchyan:2012ufa} mark a huge progress in fundamental
physics.  BICEP2 result, if confirmed, suggests an inflaton
potential with a large field amplitude, where a simple quadratic
potential $V = \frac{1}{2} M^2 \phi^2$ \cite{Linde:1983gd} is preferred.  Because the
potential needs to maintain this form up to an amplitude of 15--16
times the reduced Planck scale, the inflaton field most likely does
not participate in gauge interactions to avoid large radiative
corrections.  On the other hand, the implied mass scale $M \simeq 2\times
10^{13}$~GeV \cite{Smoot:1992td} is much larger than the observed Higgs mass of
126~GeV, hinting at a mechanism to protect a large hierarchy, such as
supersymmetry.

Once inflation is considered proven, the immediate next question is
what the inflaton is.  In particular, we need to know how inflation
ends and reheats the universe, and how the baryon asymmetry is created
after inflation, given that inflation wipes out any pre-existing
baryon asymmetry.  On both questions, it is clearly important to know
how the inflaton couples to the known particles in the Standard Model.

We show in this Letter that, being a gauge singlet, 
the inflaton naturally induces the neutrino mass.  Therefore, it is possible to identify the inflaton
with the right-handed scalar neutrino \cite{Murayama:1992ua}.  Then
leptogenesis \cite{Fukugita:1986hr} is the likely mechanism for
creating the baryon asymmetry.  Tantalizingly, the picture is
suggestive of the decay of false vacuum in the landscape
\cite{Freivogel:2014hca}, where the right-handed scalar neutrino
tunnels from a local minimum to our minimum.  If so, suppression of
scalar perturbation at low $\ell$ \cite{Bousso:2014jca} and a small
negative curvature are expected.

\paragraph{$B$-mode}
---Cosmic inflation was originally proposed to solve the flatness and
horizon problems of the big bang cosmology
\cite{Guth:1980zm,Kazanas:1980tx}.\footnote{ The exponentially
  expanding universe was also studied in Refs.~\cite{Brout:1977ix,
    Starobinsky:1980te,Sato:1980yn}), where the flatness and horizon
  problems were not discussed. }  The graceful exit problem of the
original inflation was solved by the new
inflation~\cite{Linde:1981mu,Albrecht:1982wi} where the slow-roll
inflaton drives the exponential expansion.  At the same time, it
became the dominant paradigm to generate the nearly scale-invariant,
adiabatic, and Gaussian density perturbations from the quantum
fluctuation of the inflaton field.  Its prediction has been known to
explain the data very well, including anisotropy in cosmic microwave
background radiation (CMB) and galaxy power spectrum.  However the
inflation paradigm so far lacked the definitive proof.

Primordial $B$-mode polarization of CMB is regarded as a possible
definite proof of inflation
\cite{Seljak:1996ti,Kamionkowski:1996zd,Seljak:1996gy}.  If the
expansion rate is very high during the inflationary period, gravitons
are created due to the quantum fluctuation.  Once the mode exits the
horizon, the quantum fluctuation becomes classical and the gravitons
are imprinted as primordial gravitational waves, i.e., tensor
perturbations of the space-time metric
\cite{Grishchuk:1974ny,Starobinsky:1979ty,Rubakov:1982df}.
The CMB photons acquire polarization through Thomson scatterings with
electrons on the last scattering surface because of local quadrupole
anisotropies at each point.  While density (scalar) perturbations
induce only $E$-mode polarization,
tensor perturbations induce both $E$-mode and $B$-mode polarization
patterns. Most importantly, $B$-mode polarization at small multipoles is
unlikely to be generated by other mechanisms.  The $B$-mode polarization
at large multipoles, on the other hand, is induced by gravitational
lensing effect of  large-scale structures such as clusters of galaxies 
on the way from the last scattering surface to the Earth.\footnote{{The lensing $B$-mode was recently measured
by the POLARBEAR experiment~\cite{Ade:2014afa}.}}

The recent data from BICEP2 experiment \cite{Ade:2014xna} may have
provided such a long-awaited proof of the inflation paradigm.  It
reported a detection of the $B$-mode polarization, which can be
explained by the tensor-to-scalar ratio,
$r=0.20^{+0.07}_{-0.05}$. Taken at face value, the BICEP2 results
exclude many inflation models and strongly suggest large-field
inflation models in which the inflaton field amplitude during
inflation exceeds the reduced Planck mass, $M_{Pl} =
G_N^{-1/2}/\sqrt{8\pi}\simeq 2.4 \times 10^{18}$\,GeV.  Among various
large-field inflation models, by far the simplest and therefore most
attractive one is the chaotic inflation with a simple quadratic
potential~\cite{Linde:1983gd}, which predicts $r \simeq 0.13(0.16)$
and $n_s \simeq 0.97(0.96)$ for the $e$-folding number $N_e = 60(50)$,
completely consistent with the data.  The inflaton mass is fixed to be
$M \simeq 2 \times 10^{13}$\,GeV by the normalization of density
perturbations~\cite{Smoot:1992td}.

It is worth noting that the BICEP2 results are in tension with the
Planck data \cite{Ade:2013uln} on the relative size of density perturbations on large and
small scales. This tension could be due to some unusual features in
the density perturbations such as a negative running of the scalar
spectral index~\cite{Ade:2014xna,Kobayashi:2010pz,Czerny:2014wua}, or
it may indicate the decay of a false vacuum in a landscape just before
the beginning of slow-roll
inflation~\cite{Freivogel:2014hca,Bousso:2014jca}.
The apparent tension between BICEP2 and Planck will also be partially relaxed if the true value of the tensor-to-scalar
ratio is close to the lower end of the observed range. 
We will return to this issue later in this Letter.

%
%

\paragraph{Inflaton}
---Given the inflation potentially proven, now the community should
move on to a new question: what is the inflaton?  This is a pressing
question in order to understand the subsequent cosmic history after
inflation, which depends on the coupling of the inflaton to the
standard-model particles.


On the other hand, the large hierarchy between the inflaton mass and
the Higgs mass needs to be protected by supersymmetry (see
\cite{Murayama:2000dw} for a review).  Even though the
LHC has not discovered superparticles yet, it can be hidden due to a
degenerate spectrum (see \cite{Murayama:2012jh} for a mechanism to
create a degenerate spectrum) or is simply somewhat heavier than
anticipated.  The minute supersymmetry is introduced, we need a
matter parity to avoid too-fast proton decay.  
The matter parity of the inflaton field can be either even or odd.  This
is a crucial question in order to see how the inflaton couples to the
standard-model particles.

If the inflaton field $\Phi$ has an even matter parity, its
lowest-order coupling to the standard-model particles is $W=\lambda
\Phi H_u H_d$.  Inflaton decays into Higgs fields and reheats the
Universe.\footnote{In this case, the inflaton can decay into a pair of
  gravitinos because of a possible K\"ahler potential term linear in
  the inflaton field~\cite{Kawasaki:2006gs,Kawasaki:2006hm,
    Asaka:2006bv,Dine:2006ii,Endo:2007ih,Endo:2007sz}, and tight
  cosmological constraints were obtained \cite{Nakayama:2014xca}.  See
  later discussions in this Letter how the constraints can be evaded.}
In this case, we do not see any obvious connection of the inflaton
properties to low-energy observables, nor to the origin of baryon
asymmetry.

On the other hand, if the inflaton has an odd matter parity, the
lowest-order coupling of the inflaton to the standard-model particles
is $W= h \Phi L H_u$, where $L$ and $H_u$ represent the lepton doublet
and up-type Higgs superfields, respectively, and the flavor index is
suppressed.  We expect the low-energy consequence to be $ (L
H_u)^2/M$, which is nothing but the neutrino mass.
In other words, we may say small neutrino mass is
a low-energy consequence of the inflaton.

In fact, the suggested mass of the inflaton is very close to that of
the right-handed neutrino $\approx 10^{14}$~GeV required in the seesaw
mechanism that explains small neutrino masses~\cite{Minkowski:1977sc,Yanagida:1979as,GellMann:1980vs,Yanagida:1980xy,Glashow:1979nm,Mohapatra:1979ia}.  It is
indeed a gauge-singlet.  Then we can identify the inflaton with the
right-handed scalar neutrino, as proposed some time
ago~\cite{Murayama:1992ua}.



To make the discussion more concrete, let us pick a simple model
of chaotic inflation by a quadratic potential within
supergravity~\cite{Kawasaki:2000yn,Kawasaki:2000ws}.  The
superpotential is simply the mass term
\begin{equation}
\label{MXPhi}
  W = M X \Phi,
\end{equation}
while the K\"ahler potential has a shift symmetry for the field
$\Phi \rightarrow \Phi + i c$,
\begin{equation}
  K = \frac{1}{2} (\Phi^* + \Phi)^2 + X^* X + \mbox{higher orders}.
\end{equation}
We assign the odd matter parity to $\Phi$ and $X$.

The scalar potential in supergravity reads
\begin{equation}
  \label{sugra}
  V= e^{K/M_{Pl}^2}\left(K^{I\bar{J}}
    (D_IW) (D_JW)^\dag-3\frac{|W|^2}{M_{Pl}^2}\right),
\end{equation}
with $D_I W =  \partial_I W +(\partial_J K) W/M_{Pl}^2$ and the
subscript $I$ is a label for a scalar field. 

Most importantly, the imaginary component of $\Phi$ does not appear in
the K\"ahler potential because of the shift symmetry, and therefore
the potential along ${\rm Im}[\Phi]$ remains relatively flat at super-Planckian
field values.

Let us first suppose that, during inflation, all the other fields are
stabilized at the origin. 
It is then straight-forward to work out the potential for $X$ and $\Phi$,
\begin{eqnarray}
  V &=& e^{K/M_{Pl}^2} \left(
    \left|\frac{(\Phi^* + \Phi)}{M_{Pl}^2} M X \Phi 
      + M X\right|^2 \right. \nonumber \\
  & & \left.
    + \left|\frac{X^*}{M_{Pl}^2} M X \Phi + M \Phi\right|^2 
    - 3 \left|\frac{M}{M_{Pl}} X \Phi\right|^2 \right).
\end{eqnarray}
Specializing to the imaginary term $\Phi = i \phi/\sqrt{2}$ and $X= 0$, we find
\begin{equation}
\label{Vinf}
  V = \frac{1}{2} M^2 \phi^2,
\end{equation}
a simple quadratic potential, and the correct size of density
perturbations is generated for $M \simeq 2 \times 10^{13}$\,GeV.
$X$ has the same mass for the above K\"ahler potential, but it can be stabilized at the 
origin with a positive mass of order the Hubble parameter, by adding a quartic coupling $\delta K = -
|X|^4$ in the K\"ahler potential. 

Note that the lowest-order couplings of $\Phi$ and $X$ to the
standard-model fields allowed by the odd matter parity is
\begin{equation}
  W_{\rm coupl} = h_\alpha \Phi L_\alpha H_u + \tilde{h}_\alpha X L_\alpha H_u.
\end{equation}
Then the low-energy consequence is indeed the neutrino mass
\begin{equation}
  W_{\rm eff} = \frac{1}{M} (h_\alpha L_\alpha H_u) (\tilde{h}_\beta L_\beta H_u).
\end{equation}
Therefore, it is tempting to assume that the inflaton takes part in
the origin of neutrino mass, and we hereafter identify the inflaton
field $\Phi$ as one of the right-handed neutrinos in the seesaw
mechanism.\footnote{In principle, this contribution can saturate the
  whole neutrino mass matrix with two non-zero eigenvalues.}



\paragraph{Three right-handed neutrinos}
---To be explicit, let us consider the case of three right-handed neutrinos, although 
the number of right-handed neutrinos is not restricted, {\it
  e.g.}\/, by anomaly cancellations. 
  The following arguments can be straightforwardly
  applied to a case with  right-handed neutrinos different from three.
  
The superpotential for the right-handed neutrinos $N_i$ is
\begin{equation}
  W = \frac{1}{2} M_{ij} N_i N_j + h_{i\alpha}
  N_i L_\alpha H_u,
\end{equation}
where $M_{ij}$ $(i,j=1, 2, 3)$ is the mass matrix for the
right-handed neutrinos and $h_{i\alpha}$ $(\alpha=e,\mu,\tau)$ denotes the
Yukawa coupling of the right-handed neutrino with the lepton doublet
$L_\alpha$ and the up-type Higgs $H_u$. Integrating out the heavy
right-handed neutrinos, one obtains the seesaw formula for the light
neutrino mass~\cite{Minkowski:1977sc,Yanagida:1979as,GellMann:1980vs,Yanagida:1980xy,Glashow:1979nm,Mohapatra:1979ia},
\begin{equation}
  (m_\nu)_{\alpha\beta}
  = h_{i\alpha} (M^{-1})_{ij} h_{j\beta} v_u^2,
\end{equation}
where $v_u=\langle H_u\rangle$ is the expectation value of the Higgs
field.

For successful inflation, we assume
\begin{equation}
  M_{ij} = \left(
    \begin{array}{ccc}
      m & 0 & 0\\
      0 & 0 & M\\
      0 & M & 0
    \end{array} \right)
\end{equation}
while the K\"ahler potential respects a shift symmetry for $N_3$,
\begin{equation}
  K = N_1^\dag N_1 + N_2^\dag N_2 + \frac{1}{2} (N_3^\dag + N_3)^2 + \cdots,
\end{equation}
where the dots represent higher order terms, and we suppressed $L_i$
and $H_u$ as they can be stabilized at the origin during inflation.
It is the imaginary component of $N_3$ that becomes the inflaton, and the inflaton 
potential is given by (\ref{Vinf}).

\paragraph{Reheating and Leptogenesis}
---
In order to generate a sufficiently large neutrino mass $m_\nu \approx
0.05$~eV, we need the Yukawa coupling as large as $h\sim 0.1$, where
$h$ denotes the typical value of $h_{i\alpha}$.
Then the inflaton reheats the Universe up to 
\begin{equation}
  T_{RH} \approx 
  g_*^{-\frac{1}{4}}
   \sqrt{\frac{h^2}{8\pi} M M_{Pl}}
  \gtrsim  10^{13}~\mbox{GeV},
  \label{eq:TRH}
\end{equation}
where $g_*$ counts the relativistic degrees of freedom in thermal
plasma.  For such high reheating temperature, the $e$-folding number
$N_e$ is about $60$, and the predicted values of $r$ and $n_s$ are $r
\simeq 0.13$ and $n_s \simeq 0.97$.  Also, the right-handed neutrinos
thermalize after reheating and the usual thermal leptogenesis takes
place.\footnote{ Thermal leptogenesis takes place even when the
  reheating proceeds efficiently through preheating and the subsequent
  dissipation processes.}  By keeping $m \ll M$, the $N_1$ plays the
dominant role in leptogenesis.  The CP asymmetry in its decay is given
by (see, {\it e.g.}\/, \cite{Buchmuller:2005eh})
\begin{equation}
  \epsilon_1 = \frac{1}{4\pi} 
  \frac{\Im m\sum_{\alpha,\beta}(h_{1\alpha}h_{1\beta} h^*_{2\alpha} h^*_{3\beta})}
  {\sum_{\alpha} h_{1\alpha} h^*_{1\alpha}} \frac{m}{M}\ .
\end{equation}
$m > 4 \times 10^8$~GeV is required for a successful thermal
leptogenesis \cite{Buchmuller:2002jk}.

Given the high reheating temperature Eq.~(\ref{eq:TRH}), the gravitino
is copiously produced by thermal scatterings, and its abundance is
given
by~\cite{Kawasaki:1994af,Bolz:2000fu,Pradler:2006qh,Pradler:2006hh,Rychkov:2007uq},
\begin{equation}
Y_{3/2} \simeq 2\times 10^{-9} \left(\frac{T_{RH}}{10^{13} {\rm GeV}}\right),
\end{equation}
where $Y_{3/2}$ is the ratio of the gravitino number density to the
entropy density, and we suppressed the contributions from longitudinal
mode.  Non-thermal gravitino production from inflaton decays is absent due to the matter parity~\cite{Kawasaki:2006gs}.
The gravitino decays into the Lightest Supersymmetric Particle
(LSP) at a later time.  If LSP is stable and weighs about TeV, it
exceeds the observed dark matter abundance by about four orders of
magnitude \cite{Kawasaki:1994af}.  Therefore, one possible solution is
the LSP to be as light as 100~MeV.  

This problem can be avoided also in the following ways.  One
possibility is that the gravitino is
light~\cite{Moroi:1993mb,deGouvea:1997tn}, such as in low-scale gauge
mediation models~\cite{Hamaguchi:2014sea}. The light gravitinos are
thermalized and account for the observed dark matter abundance in the
presence of mild entropy production by e.g. the lightest
messengers~\cite{Baltz:2001rq,Fujii:2002fv}.  In particular, if the
gravitino mass is lighter than about $16$\,eV~\cite{Viel:2005qj},
there is no cosmological bound on gravitinos, as their contribution to
(hot) dark matter becomes negligibly small.  If this is the case, the
NLSP would decay inside the collider detector, and the gravitino mass
can be measured in the future experiments.

Another possibility is to allow for a small matter parity
violation. Then, the LSP decays before big bang nucleosynthesis if the
gravitino mass is heavier than a few tens of TeV.  It may be tested if
the LSP is charged, or if the violation is large enough to allow for
the LSP decay within the collider detector.

If there are many singlets with an odd matter parity around the
inflaton mass scale, they also contribute to the neutrino masses, and
so, the contributions of the inflaton can be subdominant. If its
couplings are as small as $h \sim 10^{-5}$, the reheating temperature
will be of order $10^9$~GeV, greatly relaxing the gravitino
overproduction problem. In particular, the gravitino mass about
$100-1000$~TeV preferred by the anomaly mediation
\cite{Randall:1998uk,Giudice:1998xp}, the pure gravity mediation
scenario~\cite{Ibe:2011aa,Ibe:2012hu}, or the minimal split
SUSY~\cite{ArkaniHamed:2012gw,Arvanitaki:2012ps} is allowed even
without the matter parity violation. Interestingly, the observed Higgs
mass of 126~GeV can be naturally explained in this case.

\paragraph{Landscape}
---
The shift symmetry on $N_3$ is only approximate as it is explicitly broken by the
neutrino Yukawa coupling $h \sim 0.1$.  Regarding $h$ as a spurion, we expect corrections of
the form
\begin{equation}
  V = \frac{1}{2} M^2 \phi^2 \left( 1 +  c \frac{h^2 \phi^2}{M_{Pl}^2} +
    \cdots \right),
\end{equation}
with $c = {\cal O}(1)$.
Namely, the potential deviates from the simple
quadratic potential at the amplitude $\phi \sim M_{Pl}/h$.  This is
approximately the amplitude that corresponds to the $e$-folding $N
\sim 60$ to solve the flatness and horizon problems of the Universe.
It suggests the possibility that the inflation was ``just so.''



It is interesting to note that the just-so $e$-folding of $N \sim 60$
is what is expected in the landscape~\cite{Freivogel:2005vv}.  If
there are a large number of local minima in the potential, the
tunneling from the local minimum closest to ``our'' minimum should set
off the inflation.  Since a flat potential required for inflation is
not generic, the anthropic argument suggests the $e$-folding is as
small as necessary for us to exist.  Our existence requires a low
enough curvature to allow for a successful structure formation, that
leads to the lower limit $N \gtrsim 60$.  It corresponds to the
initial amplitude $\phi \sim 15 M_{Pl}$.

If taken seriously, the overall picture suggests that the right-handed
scalar neutrino tunneled from a local minimum to our minimum by the
Coleman--De Luccia mechanism.  This is a truly remarkable role for the
neutrino.  As discussed in
Refs.~\cite{Freivogel:2005vv,Bousso:2013uia}, we then expect that the
tunneling from the local minimum brings the right-handed sneutrino
where the potential is steeper than $\phi^2$, and the field starts to
roll down the potential.  Yet the field rolls very slowly because the
large negative curvature required in the Coleman--De Luccia mechanism
acts as a friction in the field equation for the right-handed
sneutrino, solving the overshoot problem.  At the same time, the
right-handed sneutrino is homogenized over many horizons solving the
initial condition problem for the chaotic inflation.  Only after the
curvature is sufficiently flattened out, the field starts to roll
faster.  Given the steeper potential at the beginning, it results in a
suppression in scalar perturbation at low $\ell$ due to its faster
motion $\delta \rho/\rho \propto V'/\dot{\phi}^2$, which ameliorates
the tension with the Planck data on the temperature
anisotropy~\cite{Freivogel:2014hca,Bousso:2014jca}.  A small negative
curvature will remain at the level of $\Omega_k \sim
10^{-4}$--$10^{-2}$.  We look forward to future precise measurements
on $\Omega_k$ from large-scale deep spectroscopic surveys such as
SuMIRe \cite{Ellis:2012rn}.

\paragraph{Conclusion}
---In this Letter, we have argued that the inflaton of chaotic
inflation with a quadratic potential suggested by the BICEP2 data
likely is a right-handed scalar neutrino.  The size of the Yukawa
coupling to generate neutrino mass violates the shift symmetry, making
the $e$-folding ``just so.''  This is what is expected in the
landscape.  The leptogenesis takes place as the reheating process
itself, or thermally after reheating.



\acknowledgments
This work was supported by  the U.S. DOE under
Contract DE-AC02-05CH11231 [HM],  by the NSF under grants PHY-1002399 and
PHY-1316783 [HM], by the Grant-in-Aid for Scientific Research (C) (No. 26400241 [HM]),
Young Scientists (B) (No.24740135) [FT],   Scientific Research on Innovative Areas (No.23104008 [FT]),
and Scientific Research (B) (No.26287039 [FT and TTY]), 
by Inoue Foundation for Science [FT], and by World Premier International Center Initiative (WPI Program), 
MEXT, Japan.

\bibliography{references}
\end{document}